\documentclass[10pt,conference]{IEEEtran}
\IEEEoverridecommandlockouts

\usepackage{amssymb}
\usepackage{amsmath, amsfonts}

\usepackage{graphicx}
\usepackage{textcomp}
\usepackage{xcolor}
\usepackage{hyperref}

\usepackage{caption}

\usepackage{xspace}
\usepackage{paralist}
\usepackage{multirow}
\usepackage{subcaption}
\usepackage{tikz}
\usepackage{courier}
\usepackage{listings} 
\usepackage{balance}
\usepackage{flushend}
\usepackage{array}
\usepackage{graphics}
\usepackage{lipsum}
\usepackage{wrapfig}
\usepackage{arydshln}
\usepackage{url}
\usepackage{alltt}
\usepackage{comment}
\usepackage{booktabs} 
\usepackage{amssymb}
\usepackage{enumitem}

\usepackage[capitalise]{cleveref}

\newcommand{\etal}{\emph{et al.}\xspace}

\NewDocumentCommand{\rangeet}
{ mO{} }{\textcolor{blue}{\textsuperscript{\textit{rangeet}}\textsf{\textbf{\small[#1]}}}}

\newcommand*{\failtopass}[0]{\mbox{$F2P$}\xspace}
\newcommand*{\oursystem}[0]{\mbox{\textsc{EvoOtter}}\xspace}
\newcommand*{\oursystemit}[0]{\mbox{\textit{E{\footnotesize VO}Otter}}\xspace}

\usepackage{cuted}

\definecolor{Gray}{gray}{0.3}
\tikzstyle{mybox} = [draw=black, very thick, rectangle, rounded corners, inner ysep=5pt, inner xsep=5pt, fill=gray!20]

\newcommand{\findings}[2]{
    \smallskip
    \noindent
    \begin{tikzpicture}
        \node [mybox] (box){%
        \centering
        \begin{minipage}{.95\columnwidth}
        \fontsize{8.8}{10}\selectfont
        \textbf{Finding #1}. #2
        \end{minipage}
        };
    \end{tikzpicture}%
}

\definecolor{darkgreen}{RGB}{0,150,0}

\begin{document}

\title{EvoOtter: Evolutionary Reproduction Test Generator}

\author{Toufique Ahmed, Jatin Ganhotra, Avraham Shinnar, and Martin Hirzel\\ IBM}

\maketitle

\begin{abstract}
Before fixing an issue, it is useful to first reproduce it by
generating a bug reproduction test (BRT).
However, generating a BRT is itself a challenging task, because issue
descriptions tend to be informal, making it difficult to determine
whether a candidate BRT indeed fails for the reason in the issue.
Prior work has attempted to tackle this problem via inference scaling,
using large language models to generate many BRTs and patches, then
using execution feedback to select and improve them.
Unfortunately, this is expensive and the feedback is unreliable.
This paper explores evolutionary programming for BRT generation
to sharpen the feedback, while enhancing evolutionary
programming to keep costs in check.
Our new approach, \oursystem, controls test execution costs via
successive halving.
Furthermore, it controls LLM costs via batched crossover for an
entire generation in a single LLM call, as well as via rule-based code
mutations, with a new fitness score tailored for BRTs.
As a result, \oursystem generates state-of-the-art quality BRTs
at the fraction of the cost of prior inference-scaling approaches to
this problem.
More broadly, this paper points at how to efficiently and effectively
combine evolutionary programming with large language models for
software engineering.
\end{abstract}


\section{Introduction}\label{sec:introduction}

A bug reproduction test~(\emph{BRT}) is a test that confirms the
presence of a bug by failing because of it~\cite{cheng_et_al_2025}.
This expectation of failure makes BRT generation different from
traditional test generation, which is expected to generate passing tests.
BRTs are important because they help with resolving issues, since they
provide feedback to a developer or agent during their bug-fixing work.
A BRT is expected to pass after the corresponding issue is resolved,
at which point it also becomes useful as a regression test to detect
recurrences of the bug.
This paper tackles the \emph{BRT generation}
problem~\cite{mundler_et_al_2024,ahmed_et_al_2024}:
\begin{description}
  \item[Given] an input $x=\langle d_\textrm{issue},c_\textrm{old}\rangle$
    consisting of an issue description $d_\textrm{issue}$ and the
    current contents of the repository, i.e., the old code
    $c_\textrm{old}$ before the issue is resolved.
  \item[Assume] a hidden oracle new code $c_\textrm{new}$ that resolves
    the issue, but is not available yet when the BRT is being generated.
  \item[Output] a generated bug reproduction test $t_\textrm{gen}$,
    such that $t_\textrm{gen}$ fails on $c_\textrm{old}$ but passes on
    $c_\textrm{new}$.
\end{description}
\noindent
In other words, a good generated bug reproduction test
$t_\textrm{gen}$ needs to be \emph{fail-to-pass}, or \failtopass for short.
What makes BRT generation challenging is the lack of reliable
feedback: since $c_\textrm{new}$ is not available, it is tricky to
improve a given BRT or to select among a set of candidate BRTs.
Another challenge is to keep costs of LLM calls and test executions low.

Several solutions have been proposed to the BRT generation
problem~\cite{ahmed_et_al_2025,ahmed_et_al_2026,cheng_et_al_2025,ehrlich_et_al_2025,khatib_mathews_nagappan_2026,wang_et_al_2024,wang_et_al_2025,wang2026icore}.
The most accurate ones tend to use inference scaling
(e.g.~\cite{wang_et_al_2024,ehrlich_et_al_2025,ahmed_et_al_2026}):
they improve their output both by sequential scaling (iterate longer on a
given candidate) and by horizontal scaling (select among many candidates).
Unfortunately, doing so relies even more on quality feedback for robustness,
while also increasing test generation costs.
On top of that, it also increases the risk of test
overfitting~\cite{smith_et_al_2015}.

This paper explores how to make inference scaling more robust via
evolutionary programming~\cite{fogel_fogel_1995}, while also lowering
the costs of that approach.
We set up a population of candidate BRTs, and the fitness of an
individual BRT measures its bug-detection capability.
Unlike most evolutionary programming algorithms, we also set up a
second population where each individual is a buggy variant of the
code.
One way to think about this is that candidate BRTs are predators and
buggy code variants are prey, such that the fittest predators catch
the most prey.
Then, the evolutionary loop uses selection and crossover to improve
the fitness of BRTs and cull out the weaker ones.
We implemented our approach in an LLM-based BRT generation workflow
called \oursystemit.

In \oursystem, the feedback for improving and selecting BRTs comes
from test execution on buggy variants of code.
To make this feedback more reliable and reduce overfitting, \oursystem
uses rule-based code mutants~\cite{jia_harman_2011}.
This has the added benefit of lowering cost: we do not use LLMs for
code mutants, but rather only reserve LLMs to work on the primary
task, which is generating and evolving BRTs.
Along the same lines, to avoid bias, we also keep the selection of
code mutants independent: the population of code mutants is chosen
randomly, and unlike BRTs, does not actively evolve.
Adamopoulos et al.\ previously used co-evolution for test generation,
using a predator/prey setup similar to
ours~\cite{adamopoulos_harman_hierons_2004}.
However, their work differs from ours in that we evolve only the
predators, not the prey; we focus on BRTs, whereas they focus on
traditional tests; and we report real experimental results, whereas
they only simulated their approach.

While our twist on evolutionary programming helps get more reliable
feedback signals, we had to introduce some additional ideas to further
reduce its costs.
\oursystem leverages successive
halving~\cite{jamieson_talwalkar_2016}: it starts from a large
population of initial candidate tests from which it selects only half
to survive at each generation, while simultaneously doubling the
population of code mutants.
That way, the feedback signal becomes stronger in each generation,
while the test execution cost per generation remains constant.
A second idea is to use batched crossover: rather than making several
LLM calls, one per crossover BRT individual, \oursystem generates the
entire generation of offspring with a single LLM call.

\oursystem achieves state-of-the-art \failtopass rates of 75.3\% on TDD-Bench-Verified and 66.3\% on SWE-Rebench using Claude-Opus-4.7. 
It also demonstrates strong performance~(77.8\%) on the SWT-Bench-Verified benchmark. 

This paper makes the following novel contributions:
\begin{itemize}
  \item \oursystem, which, besides being the first BRT generation
    workflow based on evolutionary programming, also leverages
    mutation testing and successive halving.
  \item A fitness score for BRT generation based on killing code
    mutants of the focal functions suspected to hold the bug.
  \item Batched crossover, which leverages the capabilities of modern
    LLMs while avoiding the high cost of naive LLM-based evolutionary
    programming.
\end{itemize}

Overall, this paper shows how to improve BRT generation by tailoring
evolutionary programming for modern LLMs.
We hope the ideas offered here are also useful more broadly for other
software engineering tasks.

\section{Related Work}\label{sec:relatedwork}

Our paper is the first to use evolutionary programming for BRT
generation, but there have been prior uses of evolutionary programming
for other software engineering tasks including test generation.
As mentioned in the introduction, Adamopoulos et al.\ propose
co-evolution of tests and code mutants, but differ from our work by
not creating BRTs and only conducting simulated
experiments~\cite{adamopoulos_harman_hierons_2004}.
Similarly, the popular EvoSuite uses evolutionary programming and code
mutants, but creates no BRTs~\cite{fraser_arcuri_2011}.
GenProg uses evolutionary programming for software repair instead of
software testing~\cite{legoues_et_al_2012}.
All of the above-listed works predate LLMs.
The advent of LLMs has brought about dramatic successes applying
evolutionary programming to other domains.
For instance, FunSearch evolves programs that construct evidence for
tighter bounds on hard combinatorics
problems~\cite{romeraparedes_et_al_2024},
and GEPA~(Genetic Pareto) evolves fitter LLM-based
workflows~\cite{agrawal_et_al_2025}.
However, while inspiring, neither of these is software testing related.

Our reduction in the number of LLM calls for evolutionary programming
by shrinking the prey population while growing the predator population
is derived from work in automated machine learning: DAUB performs
incremental data allocation~\cite{sabharwal_samulowitz_tesauro_2016},
and so does the successive-halving algorithm~\cite{jamieson_talwalkar_2016}.

Our paper offers a new direction for a recent and growing literature
on repository-level BRT generation.
SWE-Agent+~\cite{mundler_et_al_2024} is an adaptation of the
SWE-Agent~\cite{yang_et_al_2024} issue resolver for BRT generation.
Aegis uses execution feedback first for improving tests, then for
ranking and selection~\cite{wang_et_al_2024}.
CodeMonkeys jointly generates and improves tests with code patches for
issue resolution~\cite{ehrlich_et_al_2025}.
Agentless focuses on generating code patches and along the way also
generates BRTs to help with selection~\cite{xia_et_al_2025}.
BRT Agent generates BRTs with a ReAct action space that includes test
execution~\cite{cheng_et_al_2025}.
Otter++ uses execution feedback on $c_\textrm{old}$ to help select one
among multiple candidate tests~\cite{ahmed_et_al_2025}.
And e-Otter++ expands upon it by using feedback for both BRT
improvement and selection, and by adding execution feedback from
candidate surrogate code patches~\cite{ahmed_et_al_2026}.
AssertFlip first generates a pass-to-pass test before turning that
into a F2P test~\cite{khatib_mathews_nagappan_2026}.
Finally, iCoRe iteratively generates BRTs, then uses their results to
retrieve tests and code for context in the next round of BRT
generation~\cite{wang2026icore}.
To the best of our knowledge, none of the prior work on BRT generation
uses evolutionary programming.

\section{Methodology}\label{sec:methodology}

This section describes the overall workflow of \oursystem and its
main components (initial test generator, the selection process, and
crossover).

\subsection{Overall Workflow}
\label{sec:workflow}

\begin{figure}[t]
    \centering
    \includegraphics[width=\columnwidth]{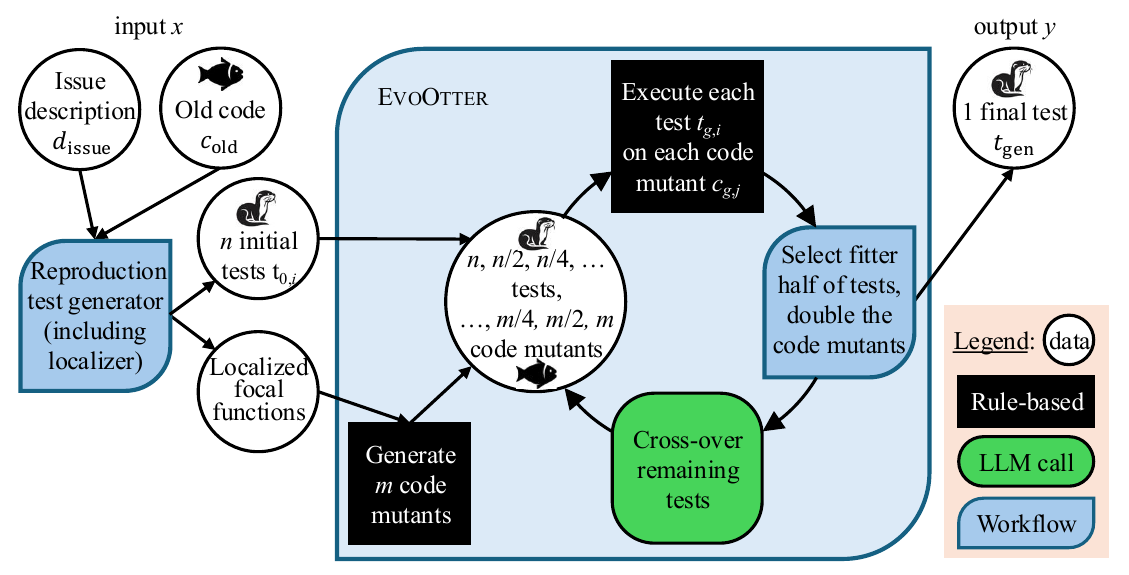}
    \caption{Overview of \oursystem workflow.}
    \label{fig:overview}
\end{figure}

\cref{fig:overview} presents the overall workflow of our approach,
\oursystem.
The input~$x$ comprises an issue description ($d_\mathrm{issue}$) and
the old code~($c_\textrm{old}$) of the repository.
Before the main part of \oursystem, we use a base reproduction test
generator to generate~$n$ distinct initial tests with
heterogeneous prompting (detailed in~\cref{sec:test-gen}).
As a by-product, this also yields a set of localized focal functions
in $c_\textrm{old}$ suspected to contain the cause of the issue.
The main part of \oursystem starts by generating a set of~$m$ code
mutants, which are modified versions of the localized focal functions.
The mutant generator uses code rewrite rules without an LLM.
The first iteration of the evolutionary loop uses all initial tests
and a small random fraction of the initial mutants.
Each iteration runs each member of the current population of candidate
tests against each member of the current population of code mutants.
The fitness score for a test is the number of mutants it killed.
We sort the tests by fitness and discard the lower half, selecting
only the fittest individuals to keep (detailed in~\cref{sec:test-selc}).
When selection leaves a single fittest test, \oursystem returns it as
the final generated test~$t_\textrm{gen}$.
Otherwise, we double the number of mutants by randomly sampling more
from the initial pre-generated set, to improve the signal for
selection in the next iteration.
Finally, we perform crossover on the test population, which uses an LLM and
also takes into account test execution logs on $c_\textrm{old}$
(detailed in~\cref{sec:crossover}).

\begin{figure}[t]
    \centering
    \includegraphics[width=\columnwidth]{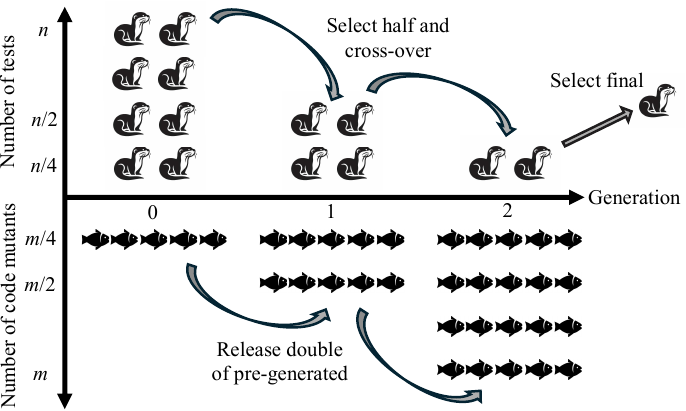}
    \captionof{figure}{Timeline.}
    \label{fig:timeline}
\end{figure}

\cref{fig:timeline} shows how \oursystem exponentially reduces the
number of tests by halving them and increases the number of mutants by
doubling them.
This approach is inspired by prior work on controlling the cost of
iterative candidate selection~\cite{sabharwal_samulowitz_tesauro_2016,jamieson_talwalkar_2016}.
Starting from an initial population of $n$ tests, this process
takes~$\log_2(n)$ loop iterations~(a.k.a.\ generations).
Each iteration executes each test on each mutant.
Thanks to the doubling/halving regime, the number of test executions
per iteration is constant at~$2m$; in the example, with $n=8$ and
$m=20$, each iteration runs~$2m=40$ test executions.
Prior work performed either selection~\cite{wang_et_al_2024, ahmed_et_al_2025, ahmed_et_al_2026, wang2026icore}, 
repair~\cite{nashid_et_al_2025, khatib_mathews_nagappan_2026, ahmed_et_al_2025, ahmed_et_al_2026}, 
or both, but in separate phases.
In contrast, \oursystem combines selection (based on mutant-killing
fitness) with repair (crossover) until it obtains a single final
test.

\subsection{Initial Test Generator}
\label{sec:test-gen}

\begin{figure}
    \centering
    \includegraphics[width=.9\columnwidth]{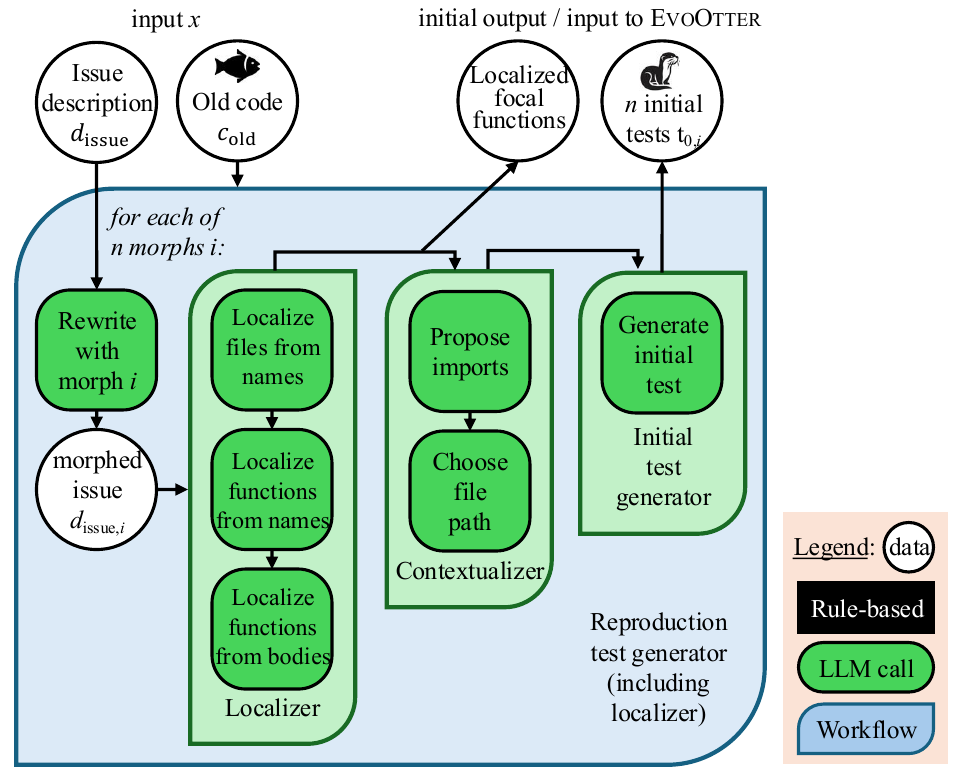}
    \caption{Initial BRT generation workflow.}
    \label{fig:basebrt}
\end{figure}

\cref{fig:basebrt} presents our base BRT generation workflow~\emph{BaseBRT}, 
which has three main components: (i) a localizer, (ii) a contextualizer, and (iii) an initial test generator.
The localizer performs three sequential LLM calls. 
First, it provides the names of all Python files in the repository and
asks the LLM to identify the 50 most relevant files.
Second, it presents these selected files along with the names of
functions they contain and asks the LLM to select relevant functions.
Finally, it refines this selection by also showing the function
bodies, enabling the LLM to further narrow down the final set of
localized focal functions.
Following localization, the contextualizer makes two sequential LLM calls. 
The first call provides the LLM with the issue description, the
localized functions (including their bodies),
and the imports extracted from the initial set of 50 files. 
It asks the LLM to identify the most relevant imports, which helps
reduce import-related difficulties later during test generation.
The second call supplies the LLM with the selected imports, the list
of candidate locations, and all previously used components, and asks
it to determine the most appropriate location within the repository
for placing the new test file.
Finally, the initial test generator constructs a complete test file
containing a fail-to-pass test case, using all information gathered
throughout the previous steps.
The generated test file is then placed in the recommended location, and a corresponding Git diff is produced. 
All prompts used in this process are provided in the supplementary material.

The workflow described above generates only one test.
However, \oursystem starts with \mbox{$n$ (usually $n=8$)} initial tests
that should ideally be both high-quality and diverse.
Prior work on inference scaling has used two different techniques for
such multi-sampling.
The most common approach is to use higher values for the
\textit{temperature} inference parameter, which increases LLM
randomness~\cite{ehrlich_et_al_2025, xia_et_al_2025, wang2026icore}.
However, with the new set of frontier models, this approach is no longer applicable. 
For example, Claude-Opus-4.7 no longer offers a temperature parameter.
Furthermore, even with earlier Claude models (e.g., Claude-Sonnet-3.7
and Claude-Sonnet-4.5), we made multiple queries at high temperature
but still obtained the same output for a certain input.
An alternative approach is heterogeneous prompting~\cite{ahmed_et_al_2026}, 
where each attempt at the task starts from a different prompt
(e.g., a different version of the issue description).
Inspired by this prior work, our initial bug reproduction test
generator uses heterogeneous prompting for generating multiple
candidates.
(\Cref{sec:crossover} also discusses a single-prompting-based approach,
which is much cheaper than heterogeneous prompting.)

Heterogeneous prompting improves diversity by systematically varying the input 
rather than relying on temperature~\cite{ahmed_et_al_2026}. 
It first applies issue description morphing, where an LLM rewrites the
original $d_\textrm{issue}$ in multiple ways,
such as standard (structured format), simple (reduced complexity and jargon), dropCode (removing misleading code), 
initTest (adding an initial test), and initPatch (adding an initial patch), 
to provide different perspectives for generation. 
It then applies context masking, where selected contextual elements, such as imports and related functions, 
are hidden from the prompt to expose the model to different subsets of information. 
We rerun the workflow shown in~\cref{fig:basebrt} with updated
information and generate multiple initial tests.
Together, these techniques encourage varied reasoning paths and increase the 
likelihood of generating at least one correct fail-to-pass test.

\subsection{Test Selection}
\label{sec:test-selc}

\begin{table}[t]
\centering
\caption{Rule-based code mutant types with examples.}
\resizebox{.95\columnwidth}{!}{%
\renewcommand{\arraystretch}{1.2}
\begin{tabular}{lll}
\toprule  
\multicolumn{1}{c}{Mutation Type} & \multicolumn{1}{c}{Before} & \multicolumn{1}{c}{After}  \\\midrule
Binary Arithmetic                 & $a + b$                    & $a - b$                    \\
Unary Arithmetic                  & $-x$                       & $+x$                       \\
Binary Relational                 & $x == y$                   & $x$ $>$ $y$                \\
Binary Logical                    & $a$ and $b$                & $a$ or $b$                 \\
Unary Logical                     & not \textit{condition}     & \textit{condition}         \\
Constant Replacement              & $x = 5$                    & $x = 0$                    \\
Return Value                      & return $x$                 & return None                \\
Assignment                        & $x = \textit{calculate}()$ & $x = \textrm{None}$        \\
Force Failure                     & -                          & \begin{tabular}[c]{@{}l@{}}raise \textit{Exception}$($\\\quad``Force failure mutant''$)$\end{tabular} \\ \bottomrule
\end{tabular}
}
\label{tbl:mutation}
\vspace{-10pt}
\end{table}

In evolutionary programming, \emph{selection} chooses fitter solutions
from the current population for the next generation~\cite{fogel_fogel_1995}.
This helps evolutionary programming gradually improve the quality of
solutions over time.
In our setup, we define the \emph{fitness} of a test as its mutation
score: the fittest tests kill the most code mutants.
\cref{tbl:mutation} presents nine different types of rule-based
mutants with examples.
In addition to traditional mutants (e.g., operator swap and constant replacement), 
it includes a mutant that just raises an exception. 
Unlike other mutants, this \emph{force failure} mutant is independent from
existing statements, so it applies even in cases where other mutants
are inapplicable.
We attempt to apply mutations to all localized functions. 
We randomly take one of the localized functions, randomly select one of the mutants, and try to apply it.
If the chosen mutant is not applicable to the function, we retry with another one.
Our rule-based mutant generator keeps going until it has
produced~$m$ unique code mutants.

Mutation testing computes the \emph{mutation score} for a test by
counting how many mutants it kills~\cite{jia_harman_2011}.
The common assumption is that the original (non-mutated) code is
correct, and that many mutations inject faults.
Therefore, traditional mutation testing starts from a test that passes
on the original code, and defines \emph{killing} a mutant simply as a
test failure, detecting the fault in the mutated code.
Unfortunately, computing the mutation score for BRTs is not that easy.
For BRTs, the original code is already incorrect, so detecting test
failures alone is insufficient.
Additionally, we do not have well-defined focal functions. 
We apply mutants to the localized functions retrieved during initial
test generation, which may have less than perfect precision or recall.
Even if our localizer can identify the focal function, it will
correspond to the old code~$c_\textrm{old}$, not the new code after
the issue is resolved.

Therefore, we propose a different definition for \emph{killing} a
mutant, resulting in a more BRT-friendly mutation score.
Instead of checking only the test outcomes, we observe changes in the
full test output logs.
In our setup, a good test should fail on~$c_\textrm{old}$ for the
reason described in~$d_\textrm{issue}$.
When a test kills a mutant, it will still fail, but for a different reason. 
Changes in the logs are a signal that the test has killed the mutant;
otherwise, it would generate the same logs.
There are challenges in computing log differences;
for example, some tests use random constants. 
To address this, we normalize whitespace and constants before computing the log differences. 
Note that the mutation score can be the same for all candidates. 
If localization fails, the mutation score may be zero for all candidates.
Conversely, in unusually simple cases, all candidates may achieve full scores. 
In such cases, test selection uses an LLM as a fallback option.
The LLM takes the issue description, tests, 
and logs as input and selects half of the candidates.
Although we could rely solely on an LLM for selection, 
doing so increases cost while reducing diversity in the population, 
thus negatively affecting the crossover phase.
Our experiments show that our mutation score better captures
execution behavior compared to an LLM, even when we give the LLM
access to the test logs.
\cref{sec:rq2} empirically compares our mutation-based 
fitness score with LLM-based selection and random selection.

\subsection{Crossover}
\label{sec:crossover}

In evolutionary programming, \emph{crossover} combines parts of
parent solutions to create new offspring with potentially better
characteristics~\cite{fogel_fogel_1995}.
By exchanging information between parents, it explores new solutions
and improves the search for an optimal result.
In our setup, we consider the set of test candidates remaining in a
given generation after selection as parents.
To generate offspring, we present the issue description, parent tests,
and logs on $c_\textrm{old}$ to an LLM.
We use a single LLM call to jointly generate the full set of diverse
offspring tests for the next generation.
We instruct the LLM to interchange, add, or remove statements between
tests to make them better.
This helps keep a diverse population while also improving individual
fitness.
We need to execute the parent tests on $c_\textrm{old}$ irrespective of the selection process, as crossover uses test logs as input.
We might get better results by making one crossover call for each candidate. 
But that would multiply the cost by the number of candidates and 
reduce our ability to control diversity as the model would not see other tests. 
Using a single call better exploits the reasoning capability of
frontier LLMs while keeping the cost in check.

Given that crossover is batched into a single LLM call, one might
wonder whether we can also generate the initial test population as a
single batch with a single LLM call.
Doing so could reduce the cost significantly, since heterogeneous
prompting is expensive, and would circumvent the lack of the
temperature parameter for temperature-based inference scaling.
On the other hand, generating new tests is harder than just
updating existing tests for crossover.
Single prompting may not match the performance of heterogeneous
prompting, because it will use the same localization and
contextualization for all candidates, whereas heterogeneous prompting
offers different components for each candidate.
We conducted some experiments with this single-prompting variant
and report results in \cref{sec:rq4}. 

To summarize, we combine selection with repair (crossover) to reach
the final test with a reduced number of LLM calls.
Ahmed \etal~\cite{ahmed_et_al_2026} applied heterogeneous prompting to generate initial candidate tests, 
but they make up to 10 LLM calls to repair each test using execution feedback before performing the selection process. 
The selection process itself also requires candidate patches, which further increases the cost. 
Moreover, with frontier models, most agents no longer generate multiple candidates.
If we apply an individual repair strategy to each test, it could give us better performance but require up to 80 LLM calls,
excluding the initial test generation and selection process. 
For many issues, fewer than 80 calls may be needed if the LLM critic determines that a test is already good and stops early. 
However, in some instances where all initial candidates fail for incorrect reasons, 
the process may still consume all 80 calls without any improvement.
In contrast, when starting from $n=8$ initial tests, \oursystem
requires at most 5 LLM calls (at most 3 for selection if the mutation
score fails and 2 for crossover, see \cref{fig:timeline}).
Empirically, we find that \oursystem only makes 3 LLM calls on average (1 for
selection and 2 for crossover), since
mutation scores usually differ across candidates. 

\section{Evaluation}
\label{sec:result}

This section describes experiments to answer the following research
questions:

\begin{itemize}[leftmargin=10pt]

\item \textbf{RQ1:} How effective is \oursystem as selector and repairer?

\item \textbf{RQ2:} How does the mutation score affect \oursystem?

\item \textbf{RQ3:} What is the contribution of crossover to \oursystem?

\item \textbf{RQ4:} How effective is single-prompt initial population generation in \oursystem?

\item \textbf{RQ5:} How does \oursystem perform overall?

\end{itemize} 

\vspace{1mm}
Prior works on repository-level BRT generation primarily evaluated on
three benchmarks: SWT-Bench-Lite~\cite{mundler_et_al_2024},
SWT-Bench-Verified~\cite{mundler_et_al_2024}, and
TDD-Bench-Verified~\cite{ahmed_et_al_2025}.
All three are derived from SWE-Bench~\cite{jimenez_et_al_2024}. 
SWT-Bench-Verified and TDD-Bench-Verified both contain subsets of instances of SWE-Bench-Verified, 
with minor differences in the evaluation harness. 
Recently, the frontier model vendor OpenAI has raised 
concerns about SWE-Bench~\cite{openai_news}, as they found some misalignment between tests and patches, 
and suggested that solutions may have been used during model training.
In this work, we primarily focus on two benchmarks: 
i)~TDD-Bench-Verified, to compare our approach with prior works,
and ii)~SWE-Rebench~\cite{badertdinov2025swe}, 
which collects new instances each month and can be used for contamination-free evaluation. 
SWE-Bench is primarily built using 12 popular and mature Python repositories, 
while SWE-Rebench collects data from more than 150 repositories. 
To avoid test framework-related complexity, 
we only focused on instances using the pytest framework. 
We executed developer-written golden tests on golden patches and checked whether the tests went from fail to pass. 
We filtered out instances where we did not observe this property. 
Finally, we ended up with 389 samples. Since SWT-Bench-Verified and TDD-Bench-Verified are similar, 
we also report the performance of \oursystem on SWT-Bench-Verified.

Following prior work, 
we use the \failtopass rate as our primary metric. 
Prior works found that \failtopass tests usually have higher coverage compared to non-\failtopass tests~\cite{ahmed_et_al_2025, ahmed_et_al_2026};
\cref{sec:discuss} briefly discusses coverage. 
Most of our experiments use Claude-Sonnet-4.5, which was released
on September 29, 2025~\cite{claude-cutoff}, 
thus allowing us to evaluate on a substantial number of instances from SWE-Rebench with pre- and post-cutoff dates. 
We also report our final results on Claude-Opus-4.7 with high reasoning to show how much we can achieve with \oursystem.

\subsection{Effectiveness of \oursystemit as Selector and Repairer (RQ1)}

\begin{table}[t]
\centering
\caption{Comparison of \oursystem with other selectors and repairers using Claude-Sonnet-4.5.}
\resizebox{.95\columnwidth}{!}{%
\renewcommand{\arraystretch}{1.2}
\begin{tabular}{l@{ }l@{ }lrr}
\toprule    
\multicolumn{1}{c}{\multirow{2}{*}{Approach}} & \multicolumn{2}{c}{\multirow{2}{*}{Skill}} & \multicolumn{2}{c}{\failtopass}                                                                                                                                                           \\
\multicolumn{1}{c}{}                          &           &                                & \multicolumn{1}{c}{\begin{tabular}[c]{@{}c@{}}TDD-Bench-Verified \\ (449 Samples)\end{tabular}} & \multicolumn{1}{c}{\begin{tabular}[c]{@{}c@{}}SWE-Rebench \\ (389 Samples)\end{tabular}} \\ \midrule
BaseBRT                                       & \multicolumn{2}{c}{Neither}                & 234 (52.1\%)                                                                                    & 147 (37.8\%)                                                                             \\
Rule-based                                    & Selection &                                & 246 (54.8\%)                                                                                    & 160 (41.1\%)                                                                             \\
LLM-based                                     & Selection &                                & 270 (60.1\%)                                                                                    & 173 (44.5\%)                                                                             \\
e-Otter                                       &           & \quad Repair                   & 284 (63.3\%)                                                                                    & 177 (45.5\%)                                                                             \\
\oursystem                                    & Selection & + Repair                       & 296 (65.9\%)                                                                                    & 202 (51.9\%)         \\ \bottomrule                                                                   
\end{tabular}
}
\label{tbl:selection}
\end{table}

Inference scaling is an effective way to improve performance. 
However, it requires a selector to choose the best candidate from a pool of generated outputs. 
Selectors can be broadly divided into two categories: (i) rule-based and (ii) LLM-based.
In rule-based selection~\cite{wang_et_al_2024,wang2026icore, ahmed_et_al_2025, ahmed_et_al_2026}, 
predefined rules are used to choose the final candidates. 
We use the Otter~\cite{ahmed_et_al_2025} selector as a baseline to compare with \oursystem. 
In this process, tests are executed on $c_\textrm{old}$ and categorized into three groups: assertion errors, 
failures due to other reasons, and runtime errors. 
Tests in the first group are prioritized, and ties are broken using a predetermined criterion, 
such as context length. If no test belongs to the first group, the remaining groups are considered sequentially.
LLMs can also be used as selectors~\cite{martinez2025dissecting,pabba2025refine}. 
This approach provides the LLM with the issue description, tests, and execution logs on 
$c_\textrm{old}$, and asks it to select the best test. 
Based on the logs, the model can determine whether a test fails for the correct reason. 
\cref{tbl:selection} shows that LLM-based selectors generally perform better than rule-based selectors.
For both selectors, we use the same initial candidates as those used for \oursystem.   
Ahmed \etal~\cite{ahmed_et_al_2026} proposed execution-based selection using surrogate code patches. 
This approach assumes that the test generator coexists with a patch generator, 
and that the patch generator also employs inference scaling. 
While this method is effective, 
it makes the overall process significantly more expensive 
(reported costs are \$2.75 and \$1.80 per instance using Claude-Sonnet-3.7 and GPT-4o, respectively). 
In this paper, we choose not to use execution feedback from surrogate code patches during the selection phase.

Execution feedback-based repair is widely used in reproduction test generation~\cite{nashid_et_al_2025, khatib_mathews_nagappan_2026, ahmed_et_al_2026}. 
We choose e-Otter as a representative baseline to compare against our approach. 
We start the repair process with the tests generated by BaseBRT using the default issue description.
\cref{tbl:selection} shows that \oursystem achieves 65.9\% and 51.9\% \failtopass 
rates on TDD-Bench-Verified and SWE-Rebench, respectively, with Claude-Sonnet-4.5, 
outperforming all individual selectors and repairers.
Note that it is always possible to repair all candidates individually and then apply a selector to achieve strong performance, 
albeit at a significantly higher cost. In contrast, \oursystem selects and batch-processes multiple candidates iteratively 
using a minimal cost (\$0.17 on average for each instance, excluding initial test generation), 
outperforming systems that rely on only one of either selection or repair.
We perform a McNemar's test~\cite{mcnemar1947note} and observe that our method 
achieves statistically significant improvements over all baselines ($p < 0.01$), except for e-Otter on TDD-Bench-Verified.

\vspace{2mm}
\findings{1}{\oursystem outperforms systems with only one of either selection or repair, and achieves 65.9\% and 51.9\% on 
TDD-Bench-Verified and SWE-Rebench, respectively, using Claude-Sonnet-4.5.}

\subsection{Influence of Mutation Score on \oursystemit (RQ2)}
\label{sec:rq2}

\begin{figure*}[t]
\centering

\begin{subfigure}{0.43\textwidth}
    \centering
    \includegraphics[width=\linewidth]{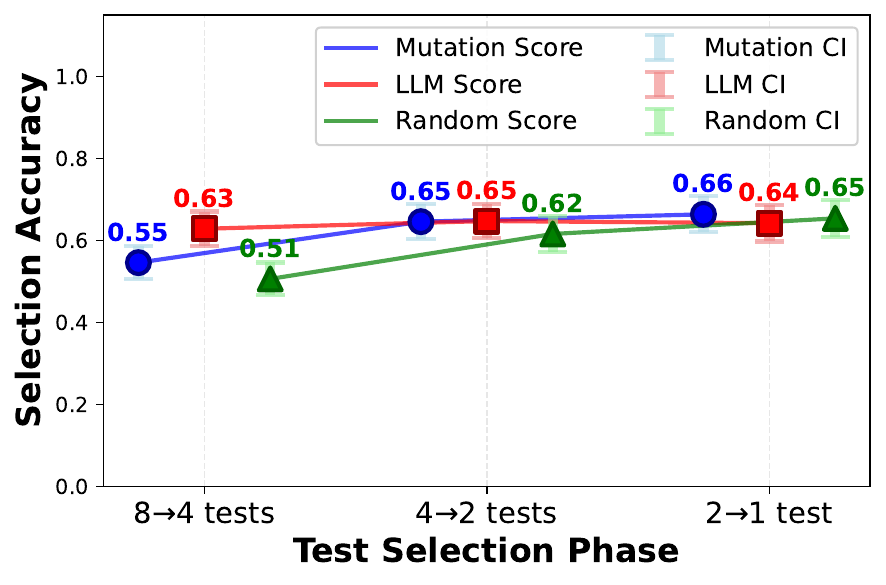}
    \caption{TDD-Bench-Verified}
\end{subfigure}
\begin{subfigure}{0.43\textwidth}
    \centering
    \includegraphics[width=\linewidth]{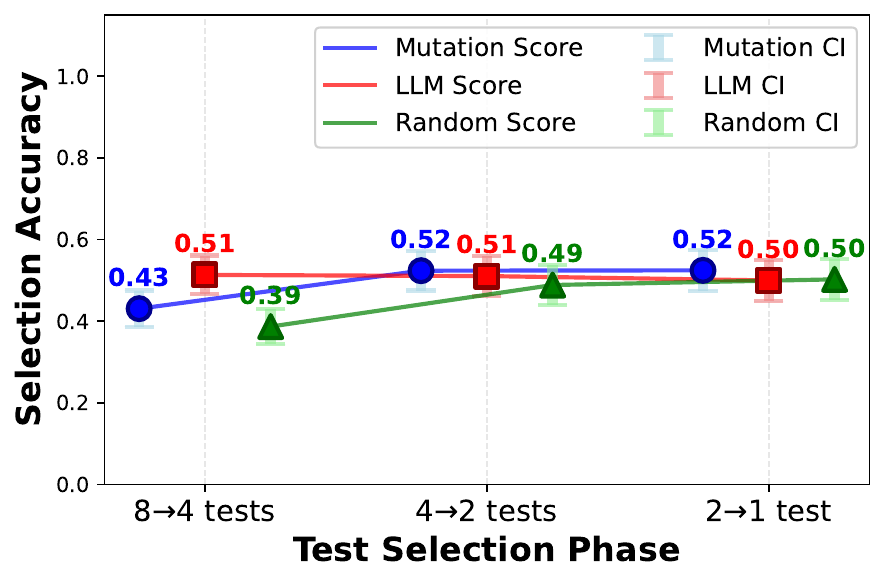}
    \caption{SWE-Rebench}
\end{subfigure}





\caption{Accuracy of selection strategies at different stages}
\label{fig:rq2}
\vspace{-10pt}
\end{figure*}

\begin{table}[t]
\centering
\caption{\oursystem with different selection and crossover mechanisms}
\resizebox{.85\columnwidth}{!}{%
\renewcommand{\arraystretch}{1.2}
\begin{tabular}{l@{ }c@{ }l@{\hspace{0.5em}}cc}
\toprule
\multicolumn{3}{c}{\multirow{2}{*}{Configuration}} & \multicolumn{2}{c}{\failtopass}                      \\
\multicolumn{3}{c}{}                               & \begin{tabular}[c]{@{}c@{}}TDD-Bench-Verified\\(449 Samples)\end{tabular} & \begin{tabular}[c]{@{}c@{}}SWE-Rebench\\(389 Samples)\end{tabular} \\ \midrule
Mutation & + & Crossover & 296 (65.9\%) & 202 (51.9\%) \\
LLM      & + & Crossover & 287 (63.9\%) & 188 (48.3\%) \\
Random   & + & Crossover & 289 (64.4\%) & 193 (49.6\%) \\
Mutation &   &           & 245 (54.6\%) & 161 (41.4\%) \\ \bottomrule
\end{tabular}
}
\label{tbl:ablation_vert}
\vspace{-10pt}
\end{table}

\begin{figure}[htbp]
    \centering

    \begin{subfigure}{0.8\columnwidth}
        \centering
        \includegraphics[width=\linewidth]{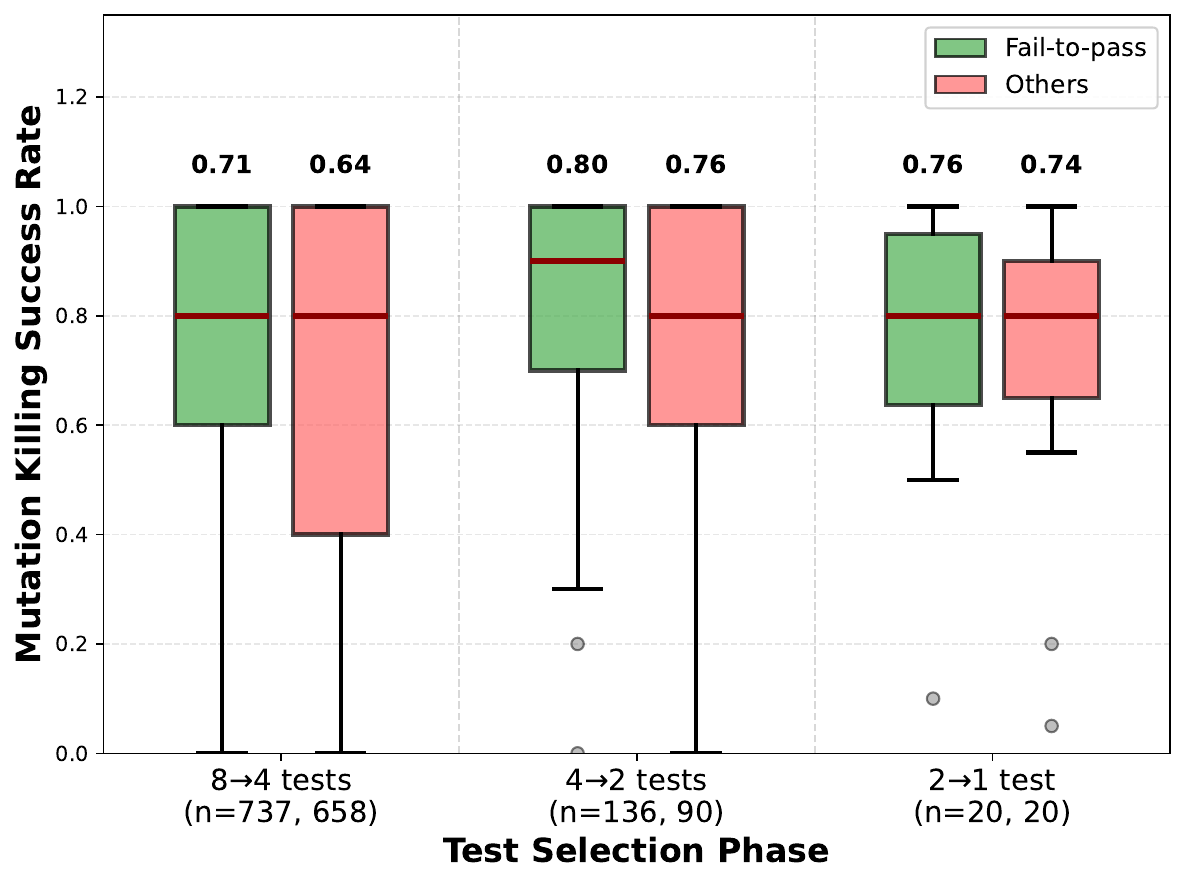}
        \caption{TDD-Bench-Verified}
        \label{fig:a}
    \end{subfigure}

    \vspace{0.2cm}

    \begin{subfigure}{0.8\columnwidth}
        \centering
        \includegraphics[width=\linewidth]{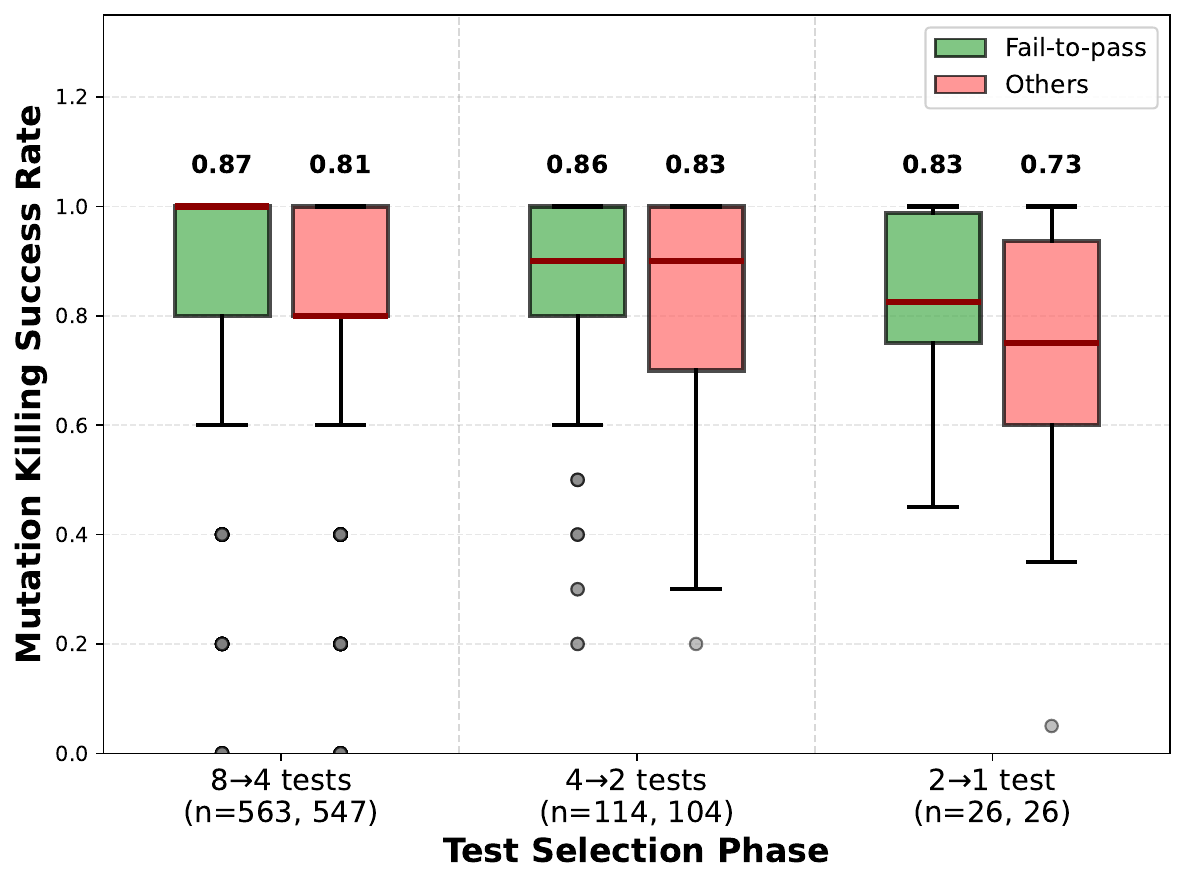}
        \caption{SWE-Rebench}
        \label{fig:b}
    \end{subfigure}

    \caption{Mutation killing ability of \failtopass tests}
    \label{fig:mutationkill}

\vspace{-10pt}    
\end{figure}

\oursystem primarily uses mutation score as its selection strategy. 
We also evaluate two additional selection strategies: LLM-based selection and random selection. 
LLM-based selection provides the LLM with the issue description, candidate tests, and execution logs on $c_\textrm{old}$,
and asks it to retain half of the population while maximizing both quality and diversity. 
Random selection samples half of the population uniformly at random for crossover. 
\cref{tbl:ablation_vert} shows that mutation-based selection achieves the best performance on both TDD-Bench-Verified and SWE-Rebench. 
Surprisingly, random selection slightly outperforms LLM-based selection when combined with crossover. 
This suggests that mutation-based and random selection benefit more from crossover than LLM-based selection. 
\cref{sec:rq3} discusses this observation further.

To further investigate the impact of selection, 
we evaluate our tests at intermediate stages against the developer-written golden patch.
Note that this evaluation does not affect the selection or crossover process; 
it is conducted solely to analyze the behavior of different components.
We consider three selection phases (e.g., $8\!\!\to\!\!4$, $4\!\!\to\!\!2$, and $2\!\!\to\!\!1$, see \cref{fig:timeline}), and in each phase, 
we compute the accuracy of the selection strategies. 
To compute accuracy, we examine how many tests are \failtopass in both the candidate population and the selected subset. 
If all selected tests are \failtopass, we assign an accuracy of 1; otherwise, we divide the number of 
selected \failtopass tests by the total number of \failtopass tests in the candidate pool.
\cref{fig:rq2} shows selection accuracy at different stages for both benchmarks. 
In the first phase, LLM-based selection performs better than the other approaches.
This is expected, as LLM-based selection has access to the execution logs on $c_\textrm{old}$, 
which can be highly beneficial. LLM critics are generally effective at identifying whether tests fail for the correct 
reasons~\cite{nashid_et_al_2025, ahmed_et_al_2026}.
However, in the subsequent phases, mutation-based and random selection begin to perform better, 
eventually outperforming LLM-based selection. 
Although the exact cause is unclear, we assume that the LLM may exhibit bias or prefer specific groups of tests,
resulting in reduced diversity among the selected tests. 
Consequently, LLM-based selection benefits the least from crossover, as reflected in the second and third selection phases.
The next section discusses the impact of crossover under different selection strategies.

\vspace{2mm}
\noindent\textbf{Does the mutation score tend to favor \failtopass tests?} If \failtopass tests and other generated 
tests exhibit similar mutant-killing capability, 
the selection process is unlikely to provide significant benefits. 
\cref{fig:mutationkill} shows that, across both benchmarks, \failtopass tests
are consistently better at killing mutants
than other generated tests. 
Although the initial selection phase contains fewer mutants, 
it provides a similar ability to distinguish high-quality tests as the later phases. 
While prioritizing \failtopass tests is not strictly required during the intermediate selection phases, 
doing so enables crossover to produce more effective offspring. 
Mutation score naturally favors \failtopass tests during selection, contributing to the improved performance of \oursystem.

\vspace{2mm}
\findings{2}{Although LLM-based selection performs best in the first phase, 
mutation-based and random selection catch up in the subsequent phases. 
Ultimately, mutation-based selection emerges as the best overall selection strategy.}

\subsection{Role of Single-Prompt Crossover in \oursystemit (RQ3)}
\label{sec:rq3}

\begin{figure*}[t]
\centering

\begin{subfigure}{0.43\textwidth}
    \centering
    \includegraphics[width=\linewidth]{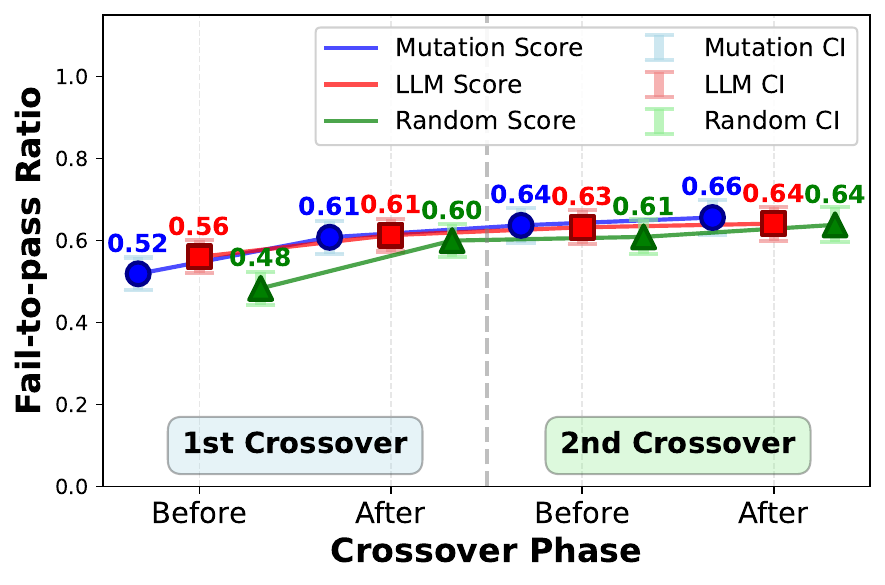}
    \caption{TDD-Bench-Verified}
\end{subfigure}
\begin{subfigure}{0.43\textwidth}
    \centering
    \includegraphics[width=\linewidth]{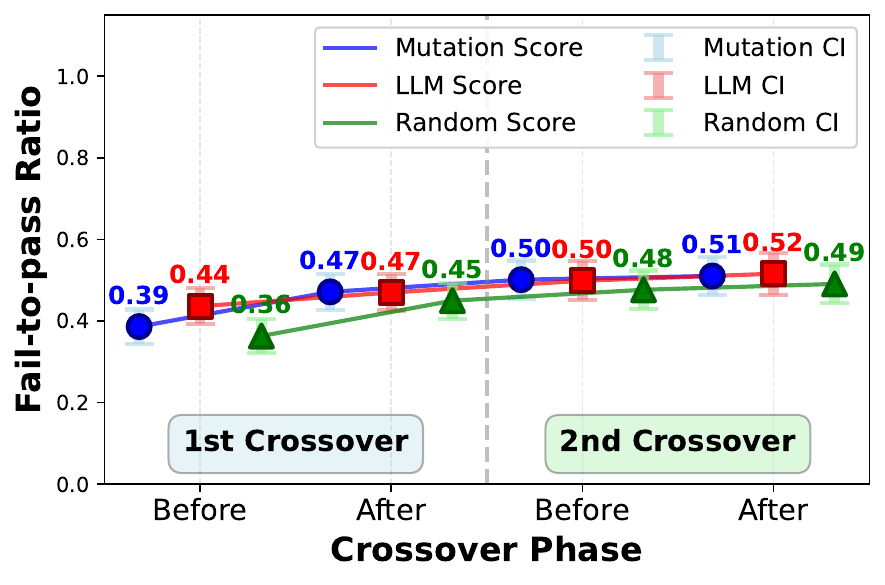}
    \caption{SWE-Rebench}
\end{subfigure}





\caption{Impact of crossover at different stages}
\label{fig:rq3}
\vspace{-10pt}
\end{figure*}

\begin{table}[t]
\centering
\caption{Evolution rate of \failtopass tests from candidate sets with no initial \failtopass tests}
\resizebox{.75\columnwidth}{!}{%
\renewcommand{\arraystretch}{1.2}

\begin{tabular}{lrrrr}
\toprule    
\multicolumn{1}{c}{\multirow{2}{*}{Category}} & \multicolumn{2}{c}{TDD-Bench-Verified}                 & \multicolumn{2}{c}{SWE-Rebench}                        \\
\multicolumn{1}{c}{}                          & \multicolumn{1}{c}{\#Sample} & \multicolumn{1}{c}{\failtopass} & \multicolumn{1}{c}{\#Sample} & \multicolumn{1}{c}{\failtopass} \\ \midrule
Mutation                                      & 287                          & 36 (12.5\%)             & 348                          & 28 (8.0\%)              \\
LLM                                           & 309                          & 19 (6.1\%)              & 326                          & 20 (6.1\%)              \\
Random                                        & 300                          & 37 (12.3\%)             & 369                          & 32 (8.7\%)     \\ \bottomrule         
\end{tabular}
}
\label{tbl:evolution}
\vspace{-10pt}
\end{table}

During crossover, we use a single LLM call to batch-process and repair all candidate tests using the issue description, 
the tests, and their corresponding execution logs on $c_\textrm{old}$. 
This is the key component of \oursystem. 
\cref{tbl:ablation_vert} shows that the choice of selection strategy has relatively lower impact when crossover is enabled. 
However, removing crossover causes the performance of mutation-based selection to drop by more than 10\% on both benchmarks, 
demonstrating the critical role of crossover in our framework.

Similar to our analysis of the selection process, 
we also examine the candidate tests at the intermediate crossover steps~(see \cref{fig:timeline}). 
Since \oursystem performs two crossovers, 
\cref{fig:rq3} reports the ratio of \failtopass tests to the 
total number of candidate tests after each crossover for both benchmarks. 
All three selection strategies benefit from crossover, 
although the improvements are larger for mutation-based and random selection. 
For example, on TDD-Bench-Verified, the \failtopass ratio increases 
by more than 9\% after the first crossover for both mutation-based and random selection,
whereas the improvement for LLM-based selection is only about 5\%. 
During the second crossover phase, the improvements are relatively small for all three strategies.
On SWE-Rebench, the \failtopass ratio after the second crossover is 0.52 for LLM-based selection, 
which is slightly higher than mutation-based selection (0.51) and random selection (0.49).
Nevertheless, mutation-based selection achieves the best overall performance. 
This is because the final performance depends on both selection and crossover. 
Although LLM-based selection produces a slightly higher \failtopass ratio after the second crossover, 
mutation-based selection retains better \failtopass tests throughout the evolutionary process, 
leading to superior overall results.

It is difficult to measure the impact of crossover when the candidate population 
already contains \failtopass tests. 
Therefore, we consider only the cases where no \failtopass tests are present 
before crossover and examine whether any new \failtopass tests emerge afterward. 
\cref{tbl:evolution} shows that, on TDD-Bench-Verified, the evolution rate is 12.5\% and 12.3\% 
for mutation-based and random selection, 
respectively, compared to only 6.1\% for LLM-based selection. 
We observe a similar trend on SWE-Rebench.
These results indicate that the candidate tests selected by mutation-based and 
random selection are more likely to evolve into \failtopass tests during crossover than those selected by the LLM-based strategy. 
We assume that this is because LLM-based selection exhibits a stronger bias 
toward a particular subset of tests, resulting in lower diversity and consequently less effective crossover.

\vspace{2mm}
\findings{3}{Crossover is the key contributor to \oursystem's performance. 
Candidates selected using mutation-based and random selection exhibit a higher 
tendency to evolve into \failtopass than those selected using the LLM-based strategy.}

\subsection{Advantages of Single-Prompt Initial Test Generation (RQ4)}
\label{sec:rq4}

\begin{table}[t]
\centering
\caption{Heterogeneous Prompting vs. Single Prompting}
\resizebox{.85\columnwidth}{!}{%
\renewcommand{\arraystretch}{1.2}

\begin{tabular}{llrr}
\toprule    
\multicolumn{1}{c}{\multirow{2}{*}{Prompting}} & \multicolumn{1}{c}{\multirow{2}{*}{Model}} & \multicolumn{2}{c}{\failtopass}                                                                                                                                                            \\
\multicolumn{1}{c}{}                           & \multicolumn{1}{c}{}                       & \multicolumn{1}{c}{\begin{tabular}[c]{@{}c@{}}TDD-Bench-Verified \\ (449 Samples)\end{tabular}} & \multicolumn{1}{c}{\begin{tabular}[c]{@{}c@{}}SWE-Rebench  \\ (389 Samples)\end{tabular}} \\ \midrule
Heterogeneous                                  & Claude-Sonnet-4.5                          & 296 (65.9\%)                                                                                    & 202 (51.9\%)                                                                              \\
Single                                         & Claude-Sonnet-4.5                          & 280 (62.4\%)                                                                                    & 194 (49.9\%)                                                                              \\
Single                                         & Claude-Opus-4.7                            & 338 (75.3\%)                                                                                    & 258 (66.3\%)     \\ \bottomrule                                                                        
\end{tabular}
}
\label{tbl:hetero-single}
\vspace{-5pt}
\end{table}

\begin{figure}[t]

    \centering
    \includegraphics[width=\linewidth]{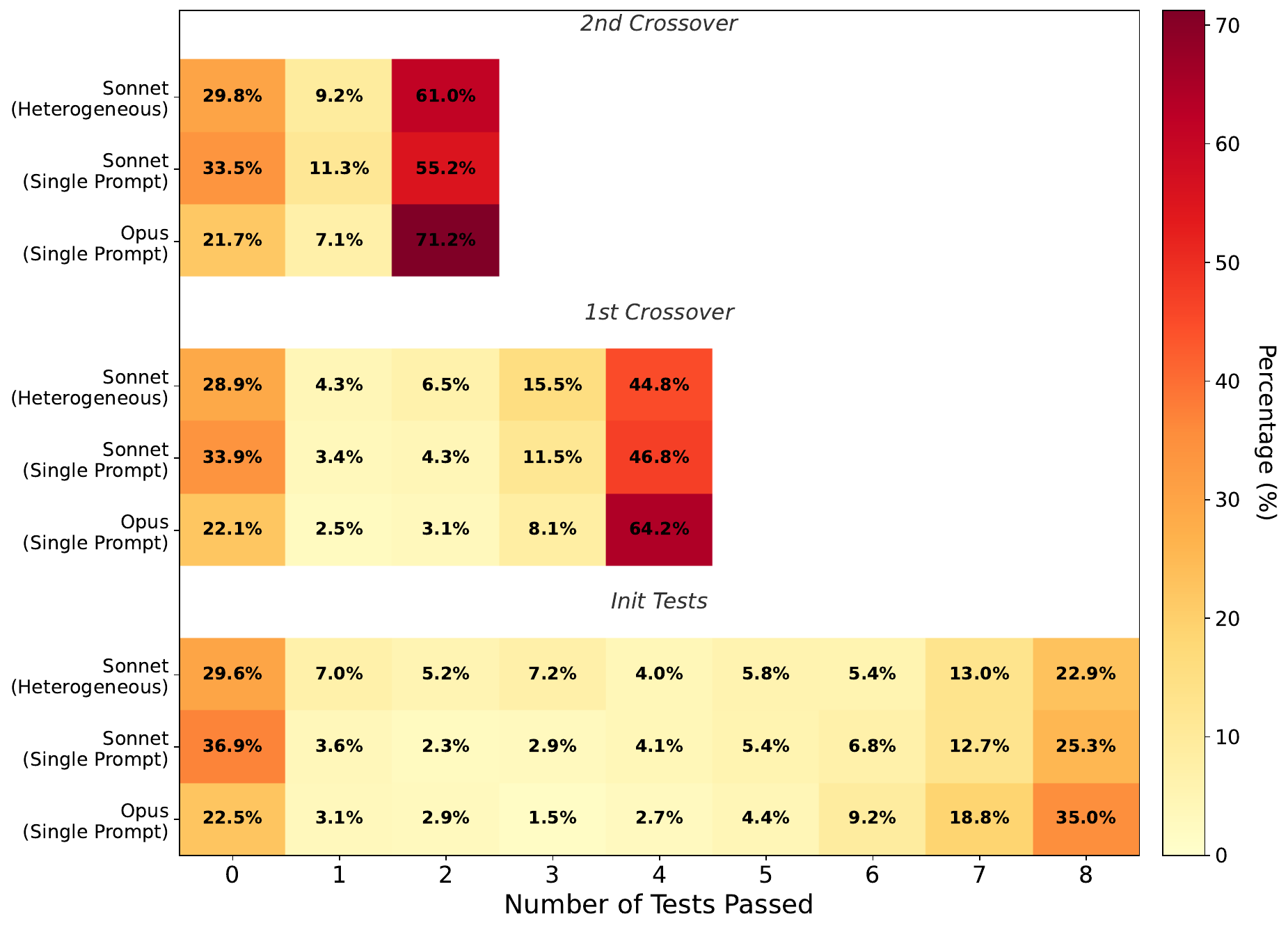}
    \caption{Distribution of \failtopass tests for heterogeneous and single prompting on TDD-Bench-Verified}
    \label{fig:heatmap}
\vspace{-10pt}    
\end{figure}

\begin{table}[t]
\centering
\caption{Estimated cost of different \oursystem variants}
\resizebox{.85\columnwidth}{!}{%
\renewcommand{\arraystretch}{1.2}

\begin{tabular}{lllr}
\toprule    
\multicolumn{1}{c}{Prompting} & \multicolumn{1}{c}{Selection} & \multicolumn{1}{c}{Model} & \multicolumn{1}{c}{Cost} \\ \midrule
Heterogeneous                 & Mutation                      & Claude-Sonnet-4.5         & \$1.12                   \\
Heterogeneous                 & LLM                           & Claude-Sonnet-4.5         & \$1.22                   \\
Single                      & Mutation                      & Claude-Sonnet-4.5         & \$0.48                   \\
Single                        & Mutation                      & Claude-Opus-4.7           & \$0.88                   \\ \bottomrule
\end{tabular}
}
\label{tbl:cost}
\vspace{-10pt}
\end{table}

\cref{sec:crossover} discussed the potential of using a single prompt 
instead of heterogeneous prompting for initial test generation. 
In this setting, we use BaseBRT for localization and contextualization without any modifications.
The only change is that we ask the LLM to generate eight distinct \failtopass test candidates 
in a single call instead of generating one test per call.
\cref{tbl:hetero-single} shows that single prompting results in a slight performance drop on both benchmarks 
(from 65.9\% to 62.4\% on TDD-Bench-Verified and from 51.9\% to 49.9\% on SWE-Rebench). 
This is expected because all candidate tests are generated from the same localization and contextualization, 
which may not always be optimal. 
Furthermore, a single LLM call is responsible for generating all eight tests,
making it more challenging to achieve the same level of diversity and quality as heterogeneous prompting,
where each call focuses on generating a single test. 
However, single prompting substantially reduces the test generation cost, from \$0.95 to \$0.20 per instance.

Despite the small performance degradation, 
the significant reduction in cost makes single prompting an attractive alternative. 
More importantly, it enables the use of stronger reasoning models, such as Claude-Opus-4.7 (with high reasoning), 
whose cost would be prohibitive under heterogeneous prompting. 
\cref{tbl:hetero-single} shows that using Claude-Opus-4.7 with single prompting achieves 75.3\% on TDD-Bench-Verified
and 66.3\% on SWE-Rebench, representing an absolute improvement of nearly 10\%-15\% over 
heterogeneous prompting with Claude-Sonnet-4.5. 
As presented in \cref{tbl:cost}, \oursystem with heterogeneous prompting and 
Claude-Sonnet-4.5 achieves a 65.9\% fail-to-pass 
rate at a cost of \$1.12 per instance, whereas single prompting with Claude-Opus-4.7 
achieves 75.3\% at a lower cost of \$0.88 per instance.

\vspace{2mm}
\noindent\textbf{Distribution of \failtopass tests under single prompting.} A potential concern with single prompting 
is the reduced diversity of the generated candidate population. 
Since all eight tests are produced in a single LLM call, 
they may exhibit similar behavior, resulting in either all tests passing or all tests failing. 
Such a lack of diversity could reduce the effectiveness of the subsequent selection and crossover stages.
\cref{fig:heatmap} shows the distribution of the number of \failtopass tests generated by \oursystem 
for both the initial population and the population after crossover under heterogeneous and single prompting.
Across all settings, we observe that for the majority of instances, either all tests pass or all tests fail. 
For example, with Claude-Opus-4.7 and single prompting, this accounts for approximately 58\% of the initial populations. 
Nevertheless, the remaining 42\% of instances contain varying numbers of \failtopass tests. 
A similar distribution is observed after crossover. 
These results suggest that tests that are generated in a single LLM call
do not necessarily share the same \failtopass behavior, allowing \oursystem's evolutionary operators to remain effective.
Furthermore, the distribution of \failtopass tests under single prompting 
is generally similar to that under heterogeneous prompting, 
suggesting that single prompting does not substantially reduce the diversity of the generated candidate population.

\vspace{1.5mm}
\findings{4}{Although replacing heterogeneous prompting with single prompting 
for initial test generation results in a slight performance drop, 
it significantly reduces the cost and enables the use of more powerful 
reasoning models to achieve better overall performance.}

\subsection{Performance of \oursystemit (RQ5)}
We compare \oursystem with prior approaches with one skill (i.e., either selection or repair) 
and analyze the contributions of its key components, including selection, crossover, and single prompting. 
Existing published and unpublished approaches employ different LLMs, making direct comparisons challenging. 
Moreover, several methods rely on temperature-based inference scaling or repeatedly querying the same prompt, 
techniques that are no longer applicable to many recent frontier models. 
A comprehensive comparison with all prior works is also impractical because several leading approaches are not open-source 
and would be prohibitively expensive to reproduce. 
Nevertheless, we compare \oursystem against e-Otter using the same underlying model, Claude-Sonnet-4.5, 
and find that \oursystem consistently outperforms it.

\cref{tbl:leaderboard} summarizes the performance of \oursystem on three benchmarks. 
On both TDD-Bench-Verified and SWE-Rebench, \oursystem achieves state-of-the-art performance. 
As discussed earlier, due to the substantial sample overlap between current SWT-Bench-Verified and TDD-Bench-Verified, 
we report results with Claude-Opus-4.7 (using single prompting) 
only to facilitate comparison with the top methods on the SWT-Bench-Verified leaderboard~\cite{swtbench_leaderboard}. 
We omit many leaderboard entries because of space limitations, 
and it is worth noting that most of the top-performing methods are unpublished and have not undergone peer review. 
In addition, the implementation details and inference costs of several of these methods are unavailable.

The performance of \oursystem could be further improved by increasing the initial population size or 
by repairing each generated test individually before the evolutionary process, 
as well as by repairing each offspring independently during crossover. 
However, these improvements would come at a substantially higher inference cost. 
In this work, our primary objective is to investigate the effectiveness of evolutionary programming 
for reproduction test generation while maintaining a low inference cost. 
Despite this focus on efficiency, \oursystem achieves competitive leaderboard performance 
and outperforms all published, peer-reviewed approaches.

\vspace{2mm}
\findings{5}{\oursystem achieves 
state-of-the-art performance on TDD-Bench-Verified (75.3\%) and 
SWE-Rebench (66.3\%) using Claude-Opus-4.7 with single prompting.
Despite its focus on cost minimization, \oursystem also performs 
strongly on the SWT-Bench-Verified leaderboard (77.8\%).}

\begin{table}[t]
\centering
\caption{Performance of \oursystem on benchmarks}
\resizebox{.98\columnwidth}{!}{%
\renewcommand{\arraystretch}{1.2}

\begin{tabular}{lllrr}
\toprule    
\multicolumn{1}{c}{\multirow{2}{*}{Benchmark}}                                              & \multicolumn{1}{c}{\multirow{2}{*}{Approach}} & \multicolumn{1}{c}{\multirow{2}{*}{Model}} & \multicolumn{2}{c}{\failtopass}                         \\
\multicolumn{1}{c}{}                                                                        & \multicolumn{1}{c}{}                          & \multicolumn{1}{c}{}                       & \multicolumn{1}{c}{Count} & \multicolumn{1}{c}{in \%} \\ \midrule
\multirow{5}{*}{\begin{tabular}[c]{@{}l@{}}TDD-Bench-Verified\\ (449 Samples)\end{tabular}} 
                                                                                            & iCoRe                                         & GPT-4o                                     & 237                       & 52.8                      \\
                                                                                            & e-Otter++                                     & Claude-Sonnet-3.7                          & 283                       & 63.0                        \\
                                                                                            & e-Otter                                       & Claude-Sonnet-4.5                          & 284                       & 63.3                      \\
                                                                                        & \textbf{EvoOtter (Heterogeneous)}                        &  \textbf{Claude-Sonnet-4.5}                          &  \textbf{296}                       &  \textbf{65.9}                      \\
                                                                                            &  \textbf{EvoOtter (Single Prompt)}                        &  \textbf{Claude-Opus-4.7}                            &  \textbf{338}                       &  \textbf{75.3}                      \\ \midrule
\multirow{6}{*}{\begin{tabular}[c]{@{}l@{}}SWT-Bench-Verified\\ (433 Samples)\end{tabular}} & AssertFlip                                    & GPT-4o                                     & 197                       & 45.5                      \\
                                                                                            & e-Otter++                                     & Claude-Sonnet-3.7                          & 269                       & 62.1                      \\
                                                                                            & ReProAgent*                                   & GPT-5-mini                                 & 302                       & 69.7                      \\
                                                                                            &  \textbf{EvoOtter (Single Prompt)}                        &  \textbf{Claude-Opus-4.7}                            &  \textbf{337}                       &  \textbf{77.8}                      \\
                                                                                            & OpenHands*                                    & GPT-5                                      & 346                       & 79.9                      \\
                                                                                            & LogicStar AI*                                 & Undisclosed                                & 364                       & 84.1                      \\ \midrule
\multirow{4}{*}{\begin{tabular}[c]{@{}l@{}}SWE-Rebench\\ (389 Samples)\end{tabular}}        & BaseBRT                                       & Claude-Sonnet-4.5                          & 147                         & 37.8                         \\
                                                                                            & e-Otter                                       & Claude-Sonnet-4.5                          & 177                          & 45.5                      \\
                                                                                            &  \textbf{EvoOtter (Heterogeneous)}                        &  \textbf{Claude-Sonnet-4.5}                          &  \textbf{202}                       &  \textbf{51.9}                      \\
                                                                                            &  \textbf{EvoOtter (Single Prompt)}                        &  \textbf{Claude-Opus-4.7}                            &  \textbf{258}                         &  \textbf{66.3}   \\ \bottomrule    
                                                                                            \multicolumn{5}{l}{* Unpublished and not peer-reviewed}                 
\end{tabular}
}
\label{tbl:leaderboard}
\end{table}

\section{Discussion}
\label{sec:discuss}

\subsection{Robustness of \oursystemit to Model Contamination}

One of the main concerns with benchmarking LLM-based software
engineering is model contamination or data memorization.
A common way to investigate this issue is to divide the evaluation instances 
into pre-cutoff and post-cutoff subsets based on the model's training cutoff date and 
compare the performance. 
M{\"u}ndler \etal~\cite{mundler_et_al_2024} conducted a similar analysis and found little difference between the two subsets. 
We perform the same analysis on SWE-Rebench, which contains instances from January 2025 to March 2026. 
Using the release date of Claude-Sonnet-4.5 (September 29, 2025) as the cutoff, 
we divide the dataset into 162 pre-cutoff instances and 227 post-cutoff instances.

Surprisingly, \cref{tbl:contamination} shows that \oursystem performs better on the post-cutoff subset than on the pre-cutoff subset, 
and this trend is consistent across all three selection strategies. 
Based on these results, it is difficult to draw a definitive conclusion about the impact of model contamination. 
One possible explanation is that the post-cutoff instances are inherently easier, 
allowing the LLM to achieve better performance. 
Nevertheless, the fact that \oursystem's performance does not degrade on the 
post-cutoff subset suggests that it may not be substantially affected by model contamination.

\begin{table}[t]
\centering
\caption{Performance of \oursystem on instances before vs.\ after Claude-Sonnet-4.5 knowledge cutoff}
\resizebox{.65\columnwidth}{!}{%
\renewcommand{\arraystretch}{1.2}
\begin{tabular}{llrrr}
\toprule    
\multicolumn{1}{c}{\multirow{2}{*}{Selection}} & \multicolumn{1}{c}{\multirow{2}{*}{Category}} & \multicolumn{1}{c}{\multirow{2}{*}{N}} & \multicolumn{2}{c}{\failtopass}                            \\
\multicolumn{1}{c}{}                           & \multicolumn{1}{c}{}                          & \multicolumn{1}{c}{}                   & \multicolumn{1}{c}{Count} & \multicolumn{1}{c}{in \%} \\ \midrule
\multirow{2}{*}{Mutation}                      & Pre-cutoff                                    & 162                                    & 78                        & 48.1                      \\
                                               & Post-cutoff                                   & 227                                    & 124                       & 54.6                      \\ \midrule
\multirow{2}{*}{LLM}                           & Pre-cutoff                                    & 162                                    & 74                        & 45.7                      \\
                                               & Post-cutoff                                   & 227                                    & 114                       & 50.2                      \\ \midrule
\multirow{2}{*}{Random}                        & Pre-cutoff                                    & 162                                    & 79                         & 48.8                         \\
                                               & Post-cutoff                                   & 227                                    & 114                        & 50.2      \\ \bottomrule                 
\end{tabular}
}
\label{tbl:contamination}
\vspace{-10pt}
\end{table}

\subsection{Coverage Evolution in \oursystemit}

The average coverage of \failtopass tests is typically above 0.90, 
whereas the average coverage of non-\failtopass generated tests is approximately 0.60~\cite{ahmed_et_al_2025}. 
Nevertheless, several non-\failtopass tests achieve coverage comparable to that of \failtopass tests. 
This raises an interesting question: besides evolving \failtopass tests, does \oursystem also improve test coverage?

\begin{figure}[t]

    \centering
    \includegraphics[width=.8\linewidth]{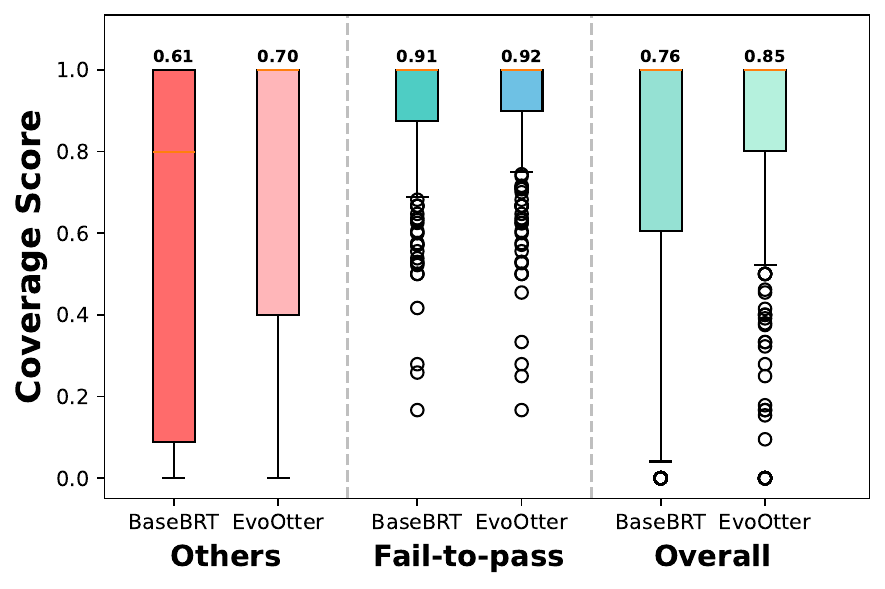}
    \caption{Evolution of coverage in generated tests}
    \label{fig:cov}
\vspace{-10pt}    
\end{figure}

Following prior work~\cite{ahmed_et_al_2025,ahmed_et_al_2026}, we compute coverage as the ratio of the number of covered deleted lines in 
$c_\textrm{old}$ and covered added lines in $c_\textrm{new}$ to the total number of added and deleted lines.
\cref{fig:cov} shows that heterogeneous prompting with 
Claude-Sonnet-4.5 achieves a substantially higher average coverage~(0.85)
than BaseBRT (0.76). When considering only \failtopass tests, the difference between the 
two methods is relatively small (0.92 vs.\ 0.91). However, for non-\failtopass tests, 
\oursystem achieves significantly higher coverage (0.70 vs.\ 0.61). 
This suggests that even when \oursystem does not generate tests with the desired property, 
it is still able to evolve tests that cover more relevant code regions.

\subsection{Detailed Analysis of the Mutants}

We use nine different mutant types in \oursystem, 
but so far have not analyzed their individual effectiveness. 
First, not every mutant type is applicable to every focal function. 
For example, the binary arithmetic mutant type does not apply if the
localized function does not contain any binary arithmetic operator.
To measure the effectiveness of each mutant type,
we try to apply mutants to the golden focal functions.
Then, using the developer-written golden test, 
we measure the mutant-killing rate.
The mutant-killing rate of the different mutant types ranges from 
72\% to 89\%. 
We also analyze the composition of the generated mutant population. 
The five operator-based mutation types collectively account for only about 25\% of all generated mutants, 
as their applicability depends on the presence of the corresponding operators in the focal function. 
In contrast, the remaining mutant types (e.g., constant replacement and force failure) 
each contribute approximately 15\%-20\% of the generated mutant population.

To evaluate the effectiveness of each mutant category in isolation, 
we group the nine mutant types into five categories, 
where all operator-based mutants are combined into a single group. 
We randomly sample 100 instances from TDD-Bench-Verified and evaluate each mutant category 
independently using heterogeneous prompting with Claude-Sonnet-4.5.
\cref{tbl:mutation-cap} shows that the default configuration, 
which uses all mutant types, achieves a 72\% \failtopass rate.
When using only a single mutant category, the \failtopass rate ranges from 66\% to 74\%. 
We note that \oursystem falls back to LLM-based selection whenever mutation scores cannot distinguish the candidate tests. 
Under the default configuration, mutation scores determine approximately 66\% of the selection decisions, 
while the remaining decisions rely on the LLM. 
However, when only a single mutant category is used, 
the proportion of LLM-based selections increases substantially. 
For example, using only the force-failure mutant category achieves a 74\% fail-to-pass rate, 
but the LLM is responsible for 61.2\% of the selection decisions. 
This suggests that comparable or even better performance may be achievable with fewer mutant categories, 
but at the expense of a higher inference cost due to increased reliance on the LLM. 
Finally, these results should be interpreted with caution, as the experiment was conducted on only 100 randomly selected instances.

\begin{table}[t]
\centering
\caption{Performance of \oursystem using different mutant categories}
\resizebox{.85\columnwidth}{!}{%
\renewcommand{\arraystretch}{1.2}

\begin{tabular}{lrrrr}
\toprule    
\multicolumn{1}{c}{\multirow{2}{*}{Mutant Type}} & \multicolumn{2}{c}{\failtopass}                      & \multicolumn{2}{c}{Select by (in \%)}                        \\ 
\multicolumn{1}{c}{}                               & \multicolumn{1}{c}{Count} & \multicolumn{1}{c}{in \%} & \multicolumn{1}{c}{Mutation Score} & \multicolumn{1}{c}{LLM} \\ \midrule
All                                                & 72                        & 72.0                        & 66.7                               & 33.3                    \\
Operators (all 5)                                  & 70                        & 70.0                        & 55.3                               & 44.7                    \\
Constant Replacement                               & 66                        & 66.0                        & 62.7                               & 37.3                    \\
Return Value                                       & 69                        & 69.0                        & 46.4                               & 53.6                    \\
Assignment                                         & 72                        & 72.0                        & 62.7                               & 37.3                    \\
Force Failure                                      & 74                        & 74.0                        & 38.8                               & 61.2                 \\ \bottomrule   
\end{tabular}
}
\label{tbl:mutation-cap}
\vspace{-10pt}
\end{table}

\section{Threats to Validity}
A limitation of our work is that it is evaluated only on Python repositories, 
and therefore the findings may not directly generalize to other programming languages. 
Nevertheless, several influential studies~\cite{jimenez_et_al_2024, mundler_et_al_2024, xia_et_al_2025} have been 
conducted under the same limitation and 
have made significant contributions to the field. 
Another important concern in SWE-bench-based evaluation is model contamination or memorization~\cite{liang2025swe}. 
To mitigate this threat, we evaluate \oursystem on SWE-Rebench, 
which contains more recent software engineering tasks, and 
observe consistent improvements across our proposed techniques. 
However, SWE-Rebench is also constructed from open-source repositories. 
Although the model is unlikely to have seen these specific tasks during training, 
it may still benefit from prior exposure to earlier versions of the corresponding repositories.
Another limitation of our study is the limited exploration of hyperparameter 
choices due to the high inference cost. 
For example, \oursystem initializes the evolutionary process with eight candidate tests.
While it would be interesting to investigate the effect of using different 
population sizes (e.g., 4 or 16 tests), such experiments are computationally expensive. 
We choose an initial population of eight because it is consistent with prior 
inference-scaling approaches, which typically generate between five and 
ten candidate tests~\cite{wang_et_al_2024, ahmed_et_al_2026, wang2026icore}. 
We expect our findings to remain largely robust to such design choices, 
although a more comprehensive sensitivity analysis is left for future work.
Finally, some of our design choices, such as single-prompt initial test generation and crossover, 
are tailored to the capabilities of recent frontier models. 
Earlier generations of LLMs may not achieve the same level of performance under these default design choices.

\section{Conclusion}

This paper introduces \oursystem, 
a simple and effective approach for generating bug reproduction tests using evolutionary programming. 
Our method uses mutation-based feedback and efficient scaling techniques to improve both accuracy and cost. 
It achieves strong \failtopass results across multiple benchmarks, showing its practical value. 
Overall, this work highlights how combining evolutionary ideas with LLMs can improve test generation in software engineering.

\bibliographystyle{IEEEtranS}  
\bibliography{bibfile}

@misc{claude-cutoff,
    url = "https://www.anthropic.com/news/claude-sonnet-4-5",
year = 2025,
}

@misc{openai_news,
    url = "https://openai.com/index/why-we-no-longer-evaluate-swe-bench-verified/",
year = 2026,
}

@misc{swtbench_leaderboard,
    url = "https://swtbench.com/?results=verified",
year = 2025,
}

@Misc{badertdinov2025swe,
  title={SWE-rebench: An Automated Pipeline for Task Collection and Decontaminated Evaluation of Software Engineering Agents},
  author={Badertdinov, Ibragim and Golubev, Alexander and Nekrashevich, Maksim and Shevtsov, Anton and Karasik, Simon and Andriushchenko, Andrei and Trofimova, Maria and Litvintseva, Daria and Yangel, Boris},
  year = 2025,
  month = may,
  url = "https://arxiv.org/abs/2505.20411" }

@Misc{liang2025swe,
  title = "The {SWE}-Bench Illusion: When State-of-the-Art {LLMs} Remember Instead of Reason",
  author = "Liang, Shanchao and Garg, Spandan and Moghaddam, Roshanak Zilouchian",
  year = 2025,
  month = jun,
  url = "https://arxiv.org/abs/2506.12286" }

@InProceedings{adamopoulos_harman_hierons_2004,
  title = "How to Overcome the Equivalent Mutant Problem and Achieve Tailored Selective Mutation Using Co-evolution",
  author = "Adamopoulos, Konstantinos and Harman, Mark and Hierons, Robert M.",
  booktitle = "Genetic and Evolutionary Computation (GECCO)",
  year = 2004,
  pages = "1338--1349",
  url = "https://doi.org/10.1007/978-3-540-24855-2_155" }

@Misc{agrawal_et_al_2025,
  title = "{GEPA}: Reflective Prompt Evolution Can Outperform Reinforcement Learning",
  author = "Agrawal, Lakshya A and Tan, Shangyin and Soylu, Dilara and Ziems, Noah and Khare, Rishi and Opsahl-Ong, Krista and Singhvi, Arnav and Shandilya, Herumb and Ryan, Michael J and Jiang, Meng and Potts, Christopher and Sen, Koushik and Dimakis, Alexandros G. and Stoica, Ion and Klein, Dan and Zaharia, Matei and Khattab, Omar",
  year = 2025,
  month = jul,
  url = "https://arxiv.org/abs/2507.19457" }

@Misc{ahmed_et_al_2024,
  title = "{TDD-Bench Verified}: Can {LLMs} Generate Tests for Issues Before They Get Resolved?",
  author = "Ahmed, Toufique and Hirzel, Martin and Pan, Rangeet and Shinnar, Avraham and Sinha, Saurabh",
  year = 2024,
  month = dec,
  url = "https://arxiv.org/abs/2412.02883" }

@InProceedings{ahmed_et_al_2025,
  title = "Otter: Generating Tests from Issues to Validate {SWE} Patches",
  author = "Ahmed, Toufique and Ganhotra, Jatin and Pan, Rangeet and Shinnar, Avraham and Sinha, Saurabh and Hirzel, Martin",
  booktitle = "International Conference on Machine Learning (ICML)",
  year = 2025,
  month = jul,
  url = "https://proceedings.mlr.press/v267/ahmed25b.html" }

@InProceedings{ahmed_et_al_2026,
  title = "Heterogeneous Prompting and Execution Feedback for {SWE} Issue Test Generation and Selection",
  author = "Ahmed, Toufique and Ganhotra, Jatin and Shinnar, Avraham and Hirzel, Martin",
  booktitle = "International Conference on Software Engineering (ICSE)",
  year = 2026,
  month = apr }

@Misc{cheng_et_al_2025,
  title = "Agentic Bug Reproduction for Effective Automated Program Repair at {Google}",
  author = "Cheng, Runxiang and Tufano, Michele and Cito, J{\"u}rgen and Cambronero, Jos{\'e} and Rondon, Pat and Wei, Renyao and Sun, Aaron and Chandra, Satish",
  year = 2025,
  month = feb,
  url = "https://arxiv.org/abs/2502.01821" }

@Misc{ehrlich_et_al_2025,
  title = "{CodeMonkeys}: Scaling Test-Time Compute for Software Engineering",
  author = "Ehrlich, Ryan and Brown, Bradley and Juravsky, Jordan and Clark, Ronald and Re, Christopher and Mirhoseini, Azalia",
  year = 2025,
  month = jan,
  url = "https://arxiv.org/abs/2501.14723" }

@InProceedings{fogel_fogel_1995,
  title = "An introduction to evolutionary programming",
  author = "Fogel, David B. and Fogel, Lawrence J.",
  booktitle = "European Conference on Artificial Evolution (AE)",
  year = 1995,
  month = sep,
  pages = "21--33",
  url = "https://doi.org/10.1007/3-540-61108-8_28" }

@InProceedings{fraser_arcuri_2011,
  title = "{EvoSuite}: automatic test suite generation for object-oriented software",
  author = "Fraser, Gordon and Arcuri, Andrea",
  booktitle = "Symposium and the European Conference on Foundations of Software Engineering (ESEC)",
  year = 2011,
  pages = "416--419",
  url = "https://doi.org/10.1145/2025113.2025179" }

@InProceedings{jamieson_talwalkar_2016,
  title = "Non-stochastic Best Arm Identification and Hyperparameter Optimization",
  author = "Jamieson, Kevin and Talwalkar, Ameet",
  booktitle = "Conference on Artificial Intelligence and Statistics (AISTATS)",
  year = 2016,
  month = may,
  pages = "240--248",
  url = "https://proceedings.mlr.press/v51/jamieson16.html" }

@Article{jia_harman_2011,
  title = "An Analysis and Survey of the Development of Mutation Testing",
  author = "Jia, Yue and Harman, Mark",
  journal = "IEEE Transactions on Software Engineering (TSE)",
  volume = 37,
  number = 5,
  year = 2011,
  pages = "649--678",
  url = "https://doi.org/10.1109/TSE.2010.62" }

@InProceedings{jimenez_et_al_2024,
  title = "{SWE-bench}: Can Language Models Resolve Real-World {GitHub} Issues?",
  author = "Jimenez, Carlos E. and Yang, John and Wettig, Alexander and Yao, Shunyu and Pei, Kexin and Press, Ofir and Narasimhan, Karthik",
  booktitle = "International Conference on Learning Representations (ICLR)",
  year = 2024,
  month = may }

@InProceedings{khatib_mathews_nagappan_2026,
  title = "{AssertFlip}: Reproducing Bugs via Inversion of {LLM}-Generated Passing Tests",
  author = "Khatib, Lara and Mathews, Noble Saji and Nagappan, Meiyappan",
  booktitle = "International Conference on Software Engineering (ICSE)",
  year = 2026,
  month = apr }

@Article{legoues_et_al_2012,
  title = "{GenProg}: A Generic Method for Automatic Software Repair",
  author = "{Le Goues}, Claire and Nguyen, ThanhVu and Forrest, Stephanie and Weimer, Westley",
  journal = "IEEE Transactions on Software Engineering (TSE)",
  volume = 38,
  number = 1,
  year = 2012,
  pages = "54--72",
  url = "https://doi.org/10.1109/TSE.2011.104" }

@InProceedings{mundler_et_al_2024,
  title = "{SWT-Bench}: Testing and Validating Real-World Bug-Fixes with Code Agents",
  author = "M{\"u}ndler, Niels and M{\"u}ller, Mark Niklas and He, Jingxuan and Vechev, Martin",
  booktitle = "Conference on Neural Information Processing Systems (NeurIPS)",
  year = 2024,
  month = dec,
  url = "https://openreview.net/forum?id=9Y8zUO11EQ" }

@Misc{nashid_et_al_2025,
  title = "{Issue2Test}: Generating Reproducing Test Cases from Issue Reports",
  author = "Nashid, Noor and Bouzenia, Islem and Pradel, Michael and Mesbah, Ali",
  year = 2025,
  month = mar,
  url = "https://arxiv.org/abs/2503.16320" }

@Article{romeraparedes_et_al_2024,
  title = "Mathematical discoveries from program search with large language models",
  author = "Romera-Paredes, Bernardino and Barekatain, Mohammadamin and Novikov, Alexander and Balog, Matej and Kumar, M. Pawan and Dupont, Emilien and Ruiz, Francisco J. R. and Ellenberg, Jordan S. and Wang, Pengming and Fawzi, Omar and Kohli, Pushmeet and Fawzi, Alhussein",
  journal = "Nature",
  year = 2024,
  month = jan,
  volume = 625,
  pages = "468--475",
  url = "https://doi.org/10.1038/s41586-023-06924-6" }

@InProceedings{sabharwal_samulowitz_tesauro_2016,
  title = "Selecting Near-Optimal Learners via Incremental Data Allocation",
  author = "Sabharwal, Ashish and Samulowitz, Horst and Tesauro, Gerald",
  booktitle = "Conference on Artificial Intelligence (AAAI)",
  year = 2016,
  month = feb,
  pages = "2007--2015",
  url = "https://www.aaai.org/ocs/index.php/AAAI/AAAI16/paper/viewPaper/12524" }

@InProceedings{smith_et_al_2015,
  author = "Smith, Edward K. and Barr, Earl T. and Le Goues, Claire and Brun, Yuriy",
  title = "Is the cure worse than the disease? Overfitting in automated program repair",
  booktitle = "Symposium on the Foundations of Software Engineering (FSE)",
  year = 2015,
  month = aug,
  url = "https://doi.org/10.1145/2786805.2786825" }

@Misc{wang_et_al_2024,
  title = "{AEGIS}: An Agent-based Framework for General Bug Reproduction from Issue Descriptions",
  author = "Wang, Xinchen and Gao, Pengfei and Meng, Xiangxin and Peng, Chao and Hu, Ruida and Lin, Yun and Gao, Cuiyun",
  year = 2024,
  month = nov,
  url = "https://arxiv.org/abs/2411.18015" }

@InProceedings{wang_et_al_2025,
  title = "Co-Evolving {LLM} Coder and Unit Tester via Reinforcement Learning",
  author = "Wang, Yinjie and Yang, Ling and Tian, Ye and Shen, Ke and Wang, Mengdi",
  booktitle = "Advances in Neural Information Processing Systems",
  volume = 38,
  year = 2025,
  month = dec,
  pages = "143630--143664",
  url = "https://proceedings.neurips.cc/paper_files/paper/2025/file/d38653cdaa8e992549e1e9e1621610d7-Paper-Conference.pdf" }

@InProceedings{xia_et_al_2025,
  title = "Demystifying {LLM}-based Software Engineering Agents",
  author = "Xia, Chunqiu Steven and Deng, Yinlin and Dunn, Soren and Zhang, Lingming",
  booktitle = "Symposium on the Foundations of Software Engineering (FSE)",
  year = 2025,
  month = jun,
  pages = "801--824",
  url = "https://doi.org/10.1145/3715754" }

@InProceedings{yang_et_al_2024,
  title = "{SWE-Agent}: Agent-computer Interfaces Enable Automated Software Engineering",
  author = "Yang, John and Jimenez, Carlos E. and Wettig, Alexander and Lieret, Kilian and Yao, Shunyu and Narasimhan, Karthik and Press, Ofir",
  booktitle = "Conference on Neural Information Processing Systems (NeurIPS)",
  year = 2024,
  month = dec,
  url = "https://proceedings.neurips.cc/paper_files/paper/2024/hash/5a7c947568c1b1328ccc5230172e1e7c-Abstract-Conference.html" }

@article{mcnemar1947note,
  title={Note on the sampling error of the difference between correlated proportions or percentages},
  author={McNemar, Quinn},
  journal={Psychometrika},
  volume={12},
  number={2},
  pages={153--157},
  year={1947},
  publisher={Springer-Verlag}
}

@Misc{wang2026icore,
  title = "{iCoRe}: An Iterative Correlation-Aware Retriever for Bug Reproduction Test Generation",
  author = "Wang, Junyi and Cao, Jialun and Liu, Zhongxin",
  year = 2026,
  month = apr,
  url = "https://arxiv.org/abs/2604.19224" }

@article{martinez2025dissecting,
  title={Dissecting the swe-bench leaderboards: Profiling submitters and architectures of llm-and agent-based repair systems},
  author={Martinez, Matias and Franch, Xavier},
  journal={arXiv preprint arXiv:2506.17208},
  year={2025}
}

@article{pabba2025refine,
  title={REFINE: Enhancing Program Repair Agents through Context-Aware Patch Refinement},
  author={Pabba, Anvith and Chen, Simin and Mathai, Alex and Chakraborty, Anindya and Ray, Baishakhi},
  journal={arXiv preprint arXiv:2510.03588},
  year={2025}
}

\end{document}